\documentclass[journal]{IEEEtran}
\usepackage{amsmath}
\usepackage{amssymb} 
\usepackage[pdftex]{graphicx}
\usepackage{cite}
\usepackage{stfloats}

\usepackage{balance}

\usepackage[hidelinks]{hyperref}

\begin{document}
\title{A Framework for Weighted-Sum Energy Efficiency Maximization in Wireless Networks}

\author{Christos~N.~Efrem, and~Athanasios~D.~Panagopoulos,~\IEEEmembership{Senior~Member,~IEEE}% <-this % stops a space
\thanks{C. N. Efrem and A. D. Panagopoulos are with the School of Electrical and Computer Engineering, National Technical University of Athens, 15780 Athens, Greece (e-mails: chefr@central.ntua.gr, thpanag@ece.ntua.gr).

This article has been accepted for publication in \textit{IEEE Wireless Communications Letters}, DOI: 10.1109/LWC.2018.2864644.  Copyright \textcopyright \ 2018 IEEE. Personal use is permitted, but republication/redistribution requires IEEE permission. See \url{http://www.ieee.org/publications_standards/publications/rights/index.html} for more information.
}}% <-this % stops a space

%\markboth{}{}

\maketitle

\begin{abstract}
Weighted-sum energy efficiency (WSEE) is a key performance metric in heterogeneous networks, where the nodes may have different energy efficiency (EE) requirements. Nevertheless, WSEE maximization is a challenging problem due to its nonconvex sum-of-ratios form. Unlike previous work, this paper presents a systematic approach to WSEE maximization under not only power constraints, but also data rate constraints, using a general SINR expression. In particular, the original problem is transformed into an equivalent form, and then a sequential convex optimization (SCO) algorithm is proposed. This algorithm is theoretically guaranteed to converge for any initial feasible point, and, under suitable constraint qualifications, achieves a Karush-Kuhn-Tucker (KKT) solution. Furthermore, we provide remarkable extensions to the proposed methodology, including systems with multiple resource blocks as well as a more general power consumption model which is not necessarily a convex function of the transmit powers. Finally, numerical analysis reveals that the proposed algorithm exhibits fast convergence, low complexity, and robustness (insensitivity to initial points).
\end{abstract}


\IEEEpeerreviewmaketitle

\section{Introduction}

\IEEEPARstart{R}{ecently}, energy efficiency (EE) maximization has become a primary issue in the design of next generation wireless networks due to economic, operational and environmental concerns. Although the network global energy efficiency (GEE), namely, the ratio between the total achievable data rate and the total power consumption, has the most meaningful interpretation as a benefit-cost ratio of the whole network, it does not contain any explicit information about the individual energy efficiencies of the links. An alternative approach in order to overcome this limitation, while maintaining high global performance, is to maximize the WSEE defined as the weighted sum of the links' energy efficiencies [1].

WSEE maximization belongs to the family of sum-of-ratios optimization problems, which are often difficult to solve. In the special case where all the ratios are in concave-convex (CC) form (assuming the case of maximization problems) and the feasible set is convex, the optimization method presented in [2] can be used to globally solve the problem. On the other hand, if at least one ratio of the sum is not in CC form and/or the feasible set is nonconvex, the optimization problem becomes more challenging. In this case, the use of standard global optimization algorithms is quite limited in practice, since they exhibit high computational complexity (generally exponential in the worst case).

An energy efficient multicell multiuser precoding technique is presented in [3], where the WSEE maximization problem is transformed into a parametrized subtractive form, and then a two-layer optimization is used to solve the problem. Later, the authors in [4] investigate the design of centralized and distributed energy-efficient coordinated beamforming in multiple-input single-output (MISO) systems with a general rate-dependent power consumption model. Furthermore, a pricing-based distributed algorithm for WSEE maximization in Ad hoc networks is given in [5]. Moreover, the authors in [6] consider the downlink of a cellular OFDMA (orthogonal frequency-division multiple-access) network with base station coordination, and propose a joint scheduling and power allocation algorithm to maximize the WSEE under maximum power constraints. Finally, the joint downlink and uplink resource allocation in time division duplex (TDD) systems with carrier aggregation is studied in [7]. 

The remainder of the paper is organized as follows. In Section II we introduce the system model and formulate the WSEE maximization problem. An optimization algorithm is developed in Section III, and then interesting extensions are reported in Section IV. Finally, simulation results are provided in Section V, while Section VI concludes the paper.

\section{System Model and Problem Formulation}
We consider a wireless network with $N$ transmitters (users), $\Lambda$ receivers, and communication bandwidth $B$. Without loss of generality, we assume that each transmitter is associated to exactly one receiver, and thus $N \ge \Lambda$. Based on [1], the signal-to-interference-plus-noise-ratio (SINR) experienced by user $i$ ($1 \le i \le N$) at its intended receiver is given by the following general expression:
\begin{equation}
{\gamma _i}({\bf{p}}) = {{{\omega _{i,i}}{p_i}} \mathord{\left/
 {\vphantom {{{\omega _{i,i}}{p_i}} {\left( {\sum\nolimits_{j \ne i} {{\omega _{j,i}}{p_j}}  + {\phi _i}{p_i} + {{\cal N}_i}} \right)}}} \right.
 \kern-\nulldelimiterspace} {\left( {\sum\nolimits_{j \ne i} {{\omega _{j,i}}{p_j}}  + {\phi _i}{p_i} + {{\cal N}_i}} \right)}}
\label{SINR}
\end{equation}
where ${\bf{p}} = {\left[ {{p_1},{p_2}, \ldots ,{p_N}} \right]^T}$ is the vector of users' transmit powers, ${{\cal N}_i}$ is the equivalent noise power, while ${\omega _{j,i}}$ and ${\phi _i}$ are \textit{non-negative parameters} that do not depend on ${\bf{p}}$ (note that the self-interference term ${\phi _i}{p_i}$ may be zero). Next, the achievable data rate and power consumption (assuming the power amplifier operates in the linear region) of the ${i^{th}}$ user are given respectively by: ${R_i}({\bf{p}}) = B\,{\log _2}\left( {1 + {\gamma _i}({\bf{p}})} \right)$ and ${P_{c,i}}({p_i}) = {\mu _i}{\kern 1pt} {p_i}\, + {P_{st,i}}$, where ${\mu _i} = {1 \mathord{\left/{\vphantom {1 {{\eta _i}}}} \right.\kern-\nulldelimiterspace} {{\eta _i}}}$, with $0 < {\eta _i} \le 1$ the power amplifier efficiency, and ${P_{st,i}} > 0$ is the static dissipated power in all other circuit blocks of the ${i^{th}}$ transmitter and its intended receiver. Moreover, the EE of user $i$ (measured in bit/Joule) is defined as follows: $E{E_i}({\bf{p}}) = {{{R_i}({\bf{p}})} \mathord{\left/{\vphantom {{{R_i}({\bf{p}})} {{P_{c,i}}({p_i})}}} \right.\kern-\nulldelimiterspace} {{P_{c,i}}({p_i})}}$. Now, we can formulate the WSEE maximization problem:
\begin{equation}
\mathop {\max }\limits_{{\bf{p}}{\kern 1pt}  \in {\kern 1pt} S} \quad \text{WSEE}({\bf{p}}) = \sum\nolimits_{i = 1}^N {{w_i}E{E_i}({\bf{p}})}
\label{WSEE_problem}
\end{equation}
with feasible set $S = \{ {\mathbf{p}} \in {\mathbb{R}^N}:{\kern 1pt} \;0 \leqslant {p_i} \leqslant P_{{\kern 1pt} i}^{\max }\;\text{and}\allowbreak\;{R_i}({\mathbf{p}}) \geqslant R_i^{\min },\;1 \leqslant i \leqslant N\}$, where ${w_i}$, $P_{{\kern 1pt} i}^{\max }$  and $R_i^{\min }$ are the priority weight, the maximum transmit power and minimum required data rate of user $i$, respectively (note that ${w_i} \geqslant 0$ and $\sum\nolimits_{i = 1}^N {{w_i}}  = 1$). It can be observed that the objective function is not in sum-of-CC-ratios form (${R_i}({\mathbf{p}})$ is not concave), and therefore the optimization method in [2] cannot be used. Nevertheless, by applying the variable transformation ${\mathbf{p}} = {2^{\mathbf{q}}}$ (${p_i} = {2^{{q_i}}}$, $1 \leqslant i \leqslant N$ with ${\mathbf{q}} = {\left[ {{q_1},{q_2}, \ldots ,{q_N}} \right]^T}$), and due to the fact that the objective is an increasing function of each user's EE, we can equivalently reformulate problem \eqref{WSEE_problem} as follows: 
\begin{equation}
\mathop {\max }\limits_{({\mathbf{q}},{\mathbf{v}}){\kern 1pt}  \in {\kern 1pt}Z} \quad f({\mathbf{v}}) = \sum\nolimits_{i = 1}^N {{w_i}{\kern 1pt} {2^{{v_i}}}} 
\label{nonconvex_problem}
\end{equation}
with feasible set $Z = \{ ({\mathbf{q}},{\mathbf{v}}) \in {\mathbb{R}^{2N}}:{\kern 1pt} \;{2^{{q_i}}} \leqslant P_{{\kern 1pt} i}^{\max },\allowbreak\;{R_i}({2^{\mathbf{q}}}) \geqslant R_i^{\min }\;\text{and}\;E{E_i}({2^{\mathbf{q}}}) \geqslant {2^{{v_i}}},\;1 \leqslant i \leqslant N\}$, where ${\mathbf{v}} = {\left[ {{v_1},{v_2}, \ldots ,{v_N}} \right]^T}$ is the vector of auxiliary variables. In addition, after some mathematical operations we get $Z = \{ ({\mathbf{q}},{\mathbf{v}}) \in {\mathbb{R}^{2N}}:{\kern 1pt} \;{2^{{q_i}}} \leqslant P_{{\kern 1pt} i}^{\max },\;{\vartheta _i}({\mathbf{q}}) \geqslant 0\allowbreak\;\text{and}\;{\varphi _i}({\mathbf{q}},{v_i}) \geqslant 0,\;1 \leqslant i \leqslant N\}$, where ${\vartheta _i}({\mathbf{q}}) = {\log _2}\left( {{{{\omega _{i,i}}} \mathord{\left/{\vphantom {{{\omega _{i,i}}} {\gamma _i^{\min }}}} \right.\kern-\nulldelimiterspace} {\gamma _i^{\min }}}} \right) + {q_i} - {\log _2}\left( {\sum\nolimits_{j \ne i} {{\omega _{j,i}}{2^{{q_j}}}}  + {\phi _i}{2^{{q_i}}} + {\mathcal{N}_i}} \right)$, with $\gamma _i^{\min } = {2^{\left( {R_i^{\min }/B} \right)}} - 1$ (${\kern 1pt} \gamma _i^{\min } \geqslant 0$, since $R_i^{\min } \geqslant 0$), and ${\varphi _i}({\mathbf{q}},{v_i}) = {R_i'}({\mathbf{q}}) - {\mu _i}{\kern 1pt} {2^{{q_i} + {v_i}}} - {P_{st,i}}{2^{{v_i}}}$, with ${R_i'}({\mathbf{q}}) = {R_i}({2^{\mathbf{q}}})$. The first and the second constraints in $Z$ are convex (the log-sum-exp function is convex [8]), whereas the third constraint is nonconvex, and $f({\mathbf{v}})$ is a \textit{strictly convex} function.

\section{WSEE Maximization Algorithm}
In the sequel, we leverage the theory of SCO, [9]-[10], in order to solve problem \eqref{nonconvex_problem}. In particular, if we have a nonconvex maximization problem $\mathcal{G}$ with objective ${g_0}({\mathbf{x}})$ and compact feasible set $\{ {\mathbf{x}} \in {\mathbb{R}^n}:\;{g_i}({\mathbf{x}}) \geqslant 0,\;1 \leqslant i \leqslant I\}$, then we can achieve a KKT solution of $\mathcal{G}$ by solving a sequence of convex maximization problems ${\{ {\widetilde {\cal G}_j}\} _{j \ge 1}}$ with objective ${\widetilde g_{0,j}}({\mathbf{x}})$, compact feasible set $\{ {\mathbf{x}} \in {\mathbb{R}^n}:\;{\widetilde g_{i,j}}({\mathbf{x}}) \geqslant 0,\allowbreak\;1 \leqslant i \leqslant I\}$, and global maximum ${\mathbf{x}}_j^ *$ (${\mathbf{x}}_0^ *$ is any feasible point of $\mathcal{G}$). Moreover, we would like to emphasize that ${g_i}({\mathbf{x}})$, ${\widetilde g_{i,j}}({\mathbf{x}})$ ($0 \leqslant i \leqslant I$ and $j \geqslant 1$) are differentiable functions that satisfy \textit{three basic properties}: 1) ${g_i}({\mathbf{x}}) \geqslant {\widetilde g_{i,j}}({\mathbf{x}}),\;\forall {\mathbf{x}} \in {\mathbb{R}^n}$, 2) ${g_i}({\mathbf{x}}_{j - 1}^ * ) = {\widetilde g_{i,j}}({\mathbf{x}}_{j - 1}^ * )$, and 3) $\nabla {g_i}({\mathbf{x}}_{j - 1}^ * ) = \nabla {\widetilde g_{i,j}}({\mathbf{x}}_{j - 1}^ * )$.

In order to lower-bound the function ${\varphi _i}({\mathbf{q}},{v_i})$ we use the following logarithmic inequality [11]:
\begin{equation}
{\log _2}(1 + \gamma ) \geqslant \alpha \,{\log _2}\gamma  + \beta ,\;\;\;\forall \gamma ,\gamma ' \geqslant 0
\end{equation}
where $\alpha  = {{\gamma '} \mathord{\left/{\vphantom {{\gamma '} {(1 + \gamma ')}}} \right.\kern-\nulldelimiterspace} {(1 + \gamma ')}}$ and $\beta  = {\log _2}(1 + \gamma ') - \alpha \,{\log _2}\gamma '$. Notice that $\alpha  \geqslant 0$, while the left-hand side and the right-hand side of inequality have equal values and first-derivatives (with respect to $\gamma$) at $\gamma  = \gamma '$. Therefore, it holds that ${R_i'}({\mathbf{q}}) \geqslant \widetilde {R_i'}({\mathbf{q}})$, with $\widetilde {R_i'}({\mathbf{q}}) = B\left[ {{\beta _i} + {\alpha _i}{{\log }_2}({\omega _{i,i}})} \right] + B{\alpha _i}\left[ {{\kern 1pt} {q_i} - {{\log }_2}\left( {\sum\nolimits_{j \ne i} {{\omega _{j,i}}{2^{{q_j}}}}  + \;{\phi _i}{2^{{q_i}}} + {\mathcal{N}_i}} \right)} \right]$, which implies that ${\varphi _i}({\mathbf{q}},{v_i}) \geqslant \widetilde {{\varphi _i}}({\mathbf{q}},{v_i})$, where $\widetilde {{\varphi _i}}({\mathbf{q}},{v_i}) = \widetilde {R_i'}({\mathbf{q}}) - {\mu _i}{\kern 1pt} {2^{{q_i} + {v_i}}} - {P_{st,i}}{2^{{v_i}}}$. Due to the convexity of the log-sum-exp function and ${2^{h({\mathbf{x}})}}$ (assuming $h({\mathbf{x}})$ is convex) [8], both $\widetilde {R_i'}({\mathbf{q}})$ and $\widetilde {{\varphi _i}}({\mathbf{q}},{v_i})$ are concave functions. Furthermore, it is known that any convex and differentiable function is lower-bounded by its first-order Taylor expansion at any point [8], and therefore we have $f({\mathbf{v}}) \geqslant f({\mathbf{v'}}) + \nabla f{({\mathbf{v'}})^T}({\mathbf{v}} - {\mathbf{v'}}) = \widetilde f({\mathbf{v}}),\allowbreak\;\forall {\mathbf{v'}} \in {\mathbb{R}^N}$ (observe that $f({\mathbf{v'}}) = \widetilde f({\mathbf{v'}})$ and $\nabla f({\mathbf{v'}}) = \nabla \widetilde f({\mathbf{v'}})$). More precisely, the affine (and thus concave) function $\widetilde f({\mathbf{v}})$ is expressed as follows:
\begin{equation}
\widetilde f({\mathbf{v}}) = \sum\nolimits_{i = 1}^N {{w_i}{\kern 1pt} {2^{v_i'}}}  + \ln (2)\sum\nolimits_{i = 1}^N {{w_i}{\kern 1pt} {2^{v_i'}}(v_i - v_i')}
\end{equation}

Consequently, we can formulate the following convex maximization problem which depends on the parameters ${\boldsymbol{\alpha }} = {\left[ {{\alpha _1},{\alpha _2}, \ldots ,{\alpha _N}} \right]^T}$, ${\boldsymbol{\beta }} = {\left[ {{\beta _1},{\beta _2}, \ldots ,{\beta _N}} \right]^T}$, and the point ${\mathbf{v'}} = {\left[ {v_1' ,v_2' , \ldots ,v_N' } \right]^T}$:
\begin{equation}
\mathop {\max }\limits_{({\mathbf{q}},{\mathbf{v}}){\kern 1pt}  \in {\kern 1pt}\Omega } \;\widetilde f({\mathbf{v}})\;\; \Leftrightarrow \;\;\mathop {\max }\limits_{({\mathbf{q}},{\mathbf{v}}){\kern 1pt}  \in {\kern 1pt}\Omega } \;\pi ({\mathbf{v}}) = \sum\nolimits_{i = 1}^N {{w_i}{\kern 1pt} {2^{v_i'}}{v_i}}
\label{convex_problem}
\end{equation}
with feasible set $\Omega  = \{ ({\mathbf{q}},{\mathbf{v}}) \in {\mathbb{R}^{2N}}:{\kern 1pt} \;{2^{{q_i}}} \leqslant P_{{\kern 1pt} i}^{\max },\allowbreak\;{\vartheta _i}({\mathbf{q}}) \geqslant 0\;\text{and}\;\widetilde {{\varphi _i}}({\mathbf{q}},{v_i}) \geqslant 0,\;1 \leqslant i \leqslant N\}$. It is noted that the two problems in \eqref{convex_problem} are equivalent, since in the second problem we omit the constant terms of the objective $\widetilde f({\mathbf{v}})$. In Algorithm 1, we provide an iterative procedure to solve problem \eqref{nonconvex_problem}, which is equivalent to the initial WSEE problem \eqref{WSEE_problem}, using the notation ${\boldsymbol{\gamma'}} = {\left[ {\gamma _1' ,\gamma _2' , \ldots ,\gamma _N' } \right]^T}$ and ${\boldsymbol{\gamma}}({\mathbf{p}}) = {\left[ {{\gamma _1}({\mathbf{p}}),{\gamma _2}({\mathbf{p}}), \ldots ,{\gamma _N}({\mathbf{p}})} \right]^T}$. 

\begin{table}[h]%[!t]%[!b]
\centering
\begin{tabular}{l}
\hline
\textbf{Algorithm 1.} WSEE Maximization                                                                                                          \\ \hline
1: Choose a sufficiently small tolerance $\varepsilon  > 0$, and a feasible point                                                                                               ${\mathbf{p}}$ \\
2: Set $\ell  = 0$, ${v_i} = {\log _2}\left( {E{E_i}({\mathbf{p}})} \right)$ for $1 \leqslant i \leqslant N$, and ${f^{(0)}} = f({\mathbf{v}})$ \\
3: \textbf{repeat} \\                                                                                                                                            
4:~~~~Compute the parameter vectors ${\boldsymbol{\alpha }}$, ${\boldsymbol{\beta }}$ with ${\boldsymbol{\gamma'}} = {\boldsymbol{\gamma }}({\mathbf{p}})$ \\
5:~~~~Solve the convex maximization problem \eqref{convex_problem} with parameters ${\boldsymbol{\alpha }}$, ${\boldsymbol{\beta }}$, \\
~~~~\space\enspace and ${\mathbf{v'}} = {\mathbf{v}}$ in order to obtain a global maximum $({{\mathbf{q}}^ * },{{\mathbf{v}}^ * })$ \\
6:~~~~Set $\ell = \ell + 1$, ${\mathbf{q}} = {{\mathbf{q}}^ * }$, ${\mathbf{v}} = {{\mathbf{v}}^ * }$, ${\mathbf{p}} = {2^{\mathbf{q}}}$, and ${f^{(\ell )}} = f({\mathbf{v}})$ \\
7: \textbf{until}                                                                                                                                                  ${{\left| {{f^{(\ell )}} - {f^{(\ell  - 1)}}} \right|} \mathord{\left/
 {\vphantom {{\left| {{f^{(\ell )}} - {f^{(\ell  - 1)}}} \right|} {\left| {{f^{(\ell  - 1)}}} \right|}}} \right.
 \kern-\nulldelimiterspace} {\left| {{f^{(\ell  - 1)}}} \right|}} < \varepsilon $ \\ \hline
\end{tabular}
\end{table}

According to [9] and [10], \textit{Algorithm 1 monotonically increases the value of the objective function $f({\mathbf{v}})$ in each iteration (i.e., ${f^{(\ell )}} \geqslant {f^{(\ell  - 1)}}$) and converges. In addition, assuming suitable constraint qualifications (e.g., Slater's condition for convex problems), the final solution $({\mathbf{q}},{\mathbf{v}})$ satisfies the KKT optimality conditions of problem \eqref{nonconvex_problem}}. It is noted that Algorithm 1 does not necessarily achieve the global optimum, since KKT conditions are only necessary, but not sufficient, for optimality in the case of nonconvex optimization problems. 

\section{Extensions to the Proposed Approach}
\subsection{Systems with Multiple Resource Blocks}
Firstly, the previous analysis can be straightforwardly extended to wireless networks with multiple ($K > 1$) resource blocks of bandwidth ${B_{RB}}$ (e.g., OFDMA systems). Based on [1], the only difference is that the QoS (quality-of-service) constraints ${B_{RB}}\sum\nolimits_{k = 1}^K {{{\log }_2}\left( {1 + \gamma _i^{\left[ k \right]}} \right)}  \geqslant R_i^{\min }$, with $\gamma _i^{\left[ k \right]} = {{\omega _{i,i}^{\left[ k \right]}{2^{q_i^{\left[ k \right]}}}} \mathord{\left/
 {\vphantom {{\omega _{i,i}^{\left[ k \right]}{2^{q_i^{\left[ k \right]}}}} {\left( {\sum\nolimits_{j \ne i} {\omega _{j,i}^{\left[ k \right]}{2^{q_j^{\left[ k \right]}}}}  + \phi _i^{\left[ k \right]}{2^{q_i^{\left[ k \right]}}} + \mathcal{N}_i^{\left[ k \right]}} \right)}}} \right.
 \kern-\nulldelimiterspace} {\left( {\sum\nolimits_{j \ne i} {\omega _{j,i}^{\left[ k \right]}{2^{q_j^{\left[ k \right]}}}}  + \phi _i^{\left[ k \right]}{2^{q_i^{\left[ k \right]}}} + \mathcal{N}_i^{\left[ k \right]}} \right)}}$, are not convex now, and they should be approximated by the convex constraints ${B_{RB}}\sum\nolimits_{k = 1}^K {\left( {\alpha _i^{\left[ k \right]}{{\log }_2}\gamma _i^{\left[ k \right]} + \beta _i^{\left[ k \right]}} \right)}  \geqslant R_i^{\min }$.

\subsection{General Power Consumption Model}
Secondly, we consider a more general rate-dependent power consumption model with non-linear power terms: 
\begin{equation}
{P_{c,i}}({\mathbf{p}}) = \sum\nolimits_{m = 1}^M {{\mu _{i,m}}{\kern 1pt} p_i^m} \, + {\xi _i}{\left( {{R_i}({\mathbf{p}})} \right)^{{\delta _i}}} + {P_{st,i}}
\end{equation}
where $M$ is the order of non-linear power terms, ${\mu _{i,m}} \geqslant 0$  measured in ${\text{W}^{1 - m}}$ (${\mu _{i,1}} = {\mu _i} = {1 \mathord{\left/{\vphantom {1 {{\eta _i}}}} \right.\kern-\nulldelimiterspace} {{\eta _i}}}$), $0 < {\delta _i} \leqslant 1$, and ${\xi _i}{\kern 1pt}  \geqslant 0$ measured in $\text{W/}{(\text{bit/s})^{{\delta _i}}}$. In conventional systems, we have $M = 1$ (absence of non-linear power terms) and ${\xi _i}{\kern 1pt}  = 0$, i.e.,  ${P_{c,i}}({p_i}) = {\mu _i}{\kern 1pt} {p_i}\, + {P_{st,i}}$. The term $\sum\nolimits_{m = 2}^M {{\mu _{i,m}}{\kern 1pt} p_i^m}$ is useful in the case of transmit signals with high peak-to-average power ratio (PAPR), and/or power amplifiers with very narrow linear region. Now, the WSEE maximization problem is formulated as follows:
\begin{equation}
\mathop {\max }\limits_{{\mathbf{p}}{\kern 1pt}  \in {\kern 1pt} S} \quad \text{WSEE}'({\mathbf{p}}) = \sum\nolimits_{i = 1}^N {{w_i}{\kern 1pt} {\kern 1pt} {\psi _i}\left( {{p_i},{R_i}({\mathbf{p}})} \right)}
\label{WSEE_problem_general}
\end{equation}
where ${\psi _i}({p_i},{\rho _i}) = {{{\rho _i}} \mathord{\left/{\vphantom {{{\rho _i}} {\left( {\sum\nolimits_{m = 1}^M {{\mu _{i,m}}{\kern 1pt} p_i^m} \, + {\xi _i}\rho _i^{{\delta _i}} + {P_{st,i}}} \right)}}} \right.\kern-\nulldelimiterspace} {\left( {\sum\nolimits_{m = 1}^M {{\mu _{i,m}}{\kern 1pt} p_i^m} \, + {\xi _i}\rho _i^{{\delta _i}} + {P_{st,i}}} \right)}}$. Notice that   ${\psi _i}({p_i},{\rho _i})$ is a strictly increasing function of ${\rho _i}$ for ${p_i},{\rho _i} \geqslant 0$, since:
\begin{equation}
\frac{{\partial {\psi _i}({p_i},{\rho _i})}}{{\partial {\rho _i}}} = \tfrac{{\sum\nolimits_{m = 1}^M {{\mu _{i,m}}{\kern 1pt} p_i^m} \, + {\xi _i}\left( {1 - {\delta _i}} \right)\rho _i^{{\delta _i}} + {P_{st,i}}}}{{{{\left( {\sum\nolimits_{m = 1}^M {{\mu _{i,m}}{\kern 1pt} p_i^m} \, + {\xi _i}\rho _i^{{\delta _i}} + {P_{st,i}}} \right)}^2}}} > 0
\end{equation}
(recall that $1 - {\delta _i} \geqslant 0$ and ${P_{st,i}} > 0$). Hence, we can rewrite problem \eqref{WSEE_problem_general} in the following form: 
\begin{equation}
\mathop {\max }\limits_{({\mathbf{p}} , {\boldsymbol{\rho }}){\kern 1pt}  \in {\kern 1pt} \Gamma } \quad \sum\nolimits_{i = 1}^N {{w_i}{\kern 1pt} {\kern 1pt} {\psi _i}({p_i},{\rho _i})}
\label{equiv_problem_general}
\end{equation}
with feasible set $\Gamma  = \{ ({\mathbf{p}} , {\boldsymbol{\rho }}) \in {\mathbb{R}^{2N}}: \;{\mathbf{p}} \in S\;\;\text{and}\;\;{R_i}({\mathbf{p}}) \geqslant {\rho _i} \geqslant 0,\;1 \leqslant i \leqslant N\}$, where ${\boldsymbol{\rho }} = {\left[ {{\rho _1},{\kern 1pt} {\kern 1pt} {\rho _2}, \ldots ,{\rho _N}} \right]^T}$ is the vector of additional variables. Using the variable transformation ${\mathbf{p}} = {2^{\mathbf{q}}}$, ${\boldsymbol{\rho }} = {2^{\mathbf{y}}}$ (${\rho _i} = {2^{{y_i}}}$, $1 \leqslant i \leqslant N$ with ${\mathbf{y}} = {\left[ {{y_1},{y_2}, \ldots ,{y_N}} \right]^T}$), and because the objective is an increasing function of each ${\psi _i}({p_i},{\rho _i})$, problem \eqref{equiv_problem_general} is equivalent to:
\begin{equation}
\mathop {\max }\limits_{({\mathbf{q}} , {\mathbf{y}}, {\mathbf{v}}){\kern 1pt}  \in {\kern 1pt} {\rm T}} \quad f({\mathbf{v}}) = \sum\nolimits_{i = 1}^N {{w_i}{\kern 1pt} {2^{{v_i}}}}
\label{nonconvex_problem_general}
\end{equation}
with feasible set ${\rm T} = \{ ({\mathbf{q}} , {\mathbf{y}}, {\mathbf{v}}) \in {\mathbb{R}^{3N}}:{\kern 1pt} \;{2^{{q_i}}} \leqslant P_{{\kern 1pt} i}^{\max },\allowbreak\;{\vartheta _i}({\mathbf{q}}) \geqslant 0,\;{R'_i}({\mathbf{q}}) \geqslant {2^{{y_i}}}\;\text{and}\;{\varepsilon _i}({q_i},{y_i},{v_i}) \leqslant 0,\;1 \leqslant i \leqslant N\}$, where ${\varepsilon _i}({q_i},{y_i},{v_i}) = \sum\nolimits_{m = 1}^M {{\mu _{i,m}}{\kern 1pt} {2^{m{\kern 1pt} {q_i} + {v_i} - {y_i}}}} \, + {\xi _i}{2^{{v_i} - {\kern 1pt} \left( {1 - {\delta _i}} \right){y_i}}} + {P_{st,i}}{2^{{v_i} - {y_i}}} - 1$ (the fourth constraint is derived from ${\psi _i}({2^{{q_i}}},{2^{{y_i}}}) \geqslant {2^{{v_i}}}$). Note that only the third constraint in ${\rm T}$ is nonconvex. Therefore, we can obtain a KKT solution for problem \eqref{nonconvex_problem_general}, which is equivalent to \eqref{WSEE_problem_general}, by solving a sequence of convex problems of the following form:
\begin{equation}
\mathop {\max }\limits_{({\mathbf{q}} , {\mathbf{y}}, {\mathbf{v}}){\kern 1pt}  \in {\kern 1pt}\Psi } \widetilde f({\mathbf{v}})\; \Leftrightarrow \;\mathop {\max }\limits_{({\mathbf{q}} , {\mathbf{y}}, {\mathbf{v}}){\kern 1pt}  \in {\kern 1pt}\Psi } \pi ({\mathbf{v}}) = \sum\nolimits_{i = 1}^N {{w_i}{\kern 1pt} {2^{v_i' }}{v_i}}
\end{equation}
with feasible set $\Psi  = \{ ({\mathbf{q}} , {\mathbf{y}}, {\mathbf{v}}) \in {\mathbb{R}^{3N}}:{\kern 1pt} \;{2^{{q_i}}} \leqslant P_{{\kern 1pt} i}^{\max },\allowbreak\;{\vartheta _i}({\mathbf{q}}) \geqslant 0,\;\widetilde {R_i'}({\mathbf{q}}) \geqslant {2^{{y_i}}}\;\text{and}\;{\varepsilon _i}({q_i},{y_i},{v_i}) \leqslant 0,\;1 \leqslant i \leqslant N\}$.

\section{Numerical Results}
Consider a relay-assisted multiple-input multiple-output (MIMO) network, where $N$ transmitters communicate with $N$ receivers through a single-antenna amplify-and-forward relay (receiver $i$ is the intended receiver of transmitter $i$). We denote by ${L_T}$, ${L_R}$ the number of antennas at each transmitter and receiver, respectively. Moreover, ${{\mathbf{b}}_i}$ (with $\left\| {{{\mathbf{b}}_i}} \right\| = 1$) is the ${L_T} \times 1$ beamforming vector of transmitter $i$ (assume that ${p_i}$ is equally divided between the transmit antennas, i.e., ${{\mathbf{b}}_i} = \left( {{1 \mathord{\left/{\vphantom {1 {\sqrt {{L_T}} }}} \right.\kern-\nulldelimiterspace} {\sqrt {{L_T}} }}} \right){{\mathbf{1}}_{{L_T} \times 1}}$, with ${{\mathbf{1}}_{{L_T} \times 1}}$ the ${L_T} \times 1$ vector of ones), ${{\mathbf{h}}_i}$ is the $1 \times {L_T}$ channel vector from transmitter $i$ to the relay, ${{\mathbf{g}}_i}$ is the ${L_R} \times 1$ channel vector from the relay to receiver $i$, and ${{\mathbf{c}}_i}$ is the ${L_R} \times 1$ combining vector of receiver $i$. Also, suppose the receivers perform maximum-ratio combining (MRC), i.e., ${{\mathbf{c}}_i} = {{\mathbf{g}}_i}{{\mathbf{h}}_i}{{\mathbf{b}}_i}$. The received signal at the relay is given by ${x_r} = \sum\nolimits_{j = 1}^N {\sqrt {{p_j}} {\kern 1pt} {{\mathbf{h}}_j}{{\mathbf{b}}_j}{s_j}}  + {n_r}$, where ${s_j}$ is the information symbol of transmitter $j$ ($E\{ {s_j}\}  = 0$, $E\{ |{s_j}{|^2}\}  = 1$), and ${n_r} \sim \mathcal{C}\mathcal{N}\left( {0,\sigma _r^2} \right)$ is the relay thermal noise. Thus, the total input power at the relay is ${P_{r,in}} = \sum\nolimits_{j = 1}^N {{p_j}{{\left| {{{\mathbf{h}}_j}{{\mathbf{b}}_j}} \right|}^2}}  + \sigma _r^2$. Then, the received signal at the relay is normalized by $\sqrt {{P_{r,in}}}$, before being amplified by a factor $\sqrt {{P_r}}$ (${P_r}$ is the relay transmit power) and forwarded to the receivers, in order to ensure that the relay power amplifier operates within the linear region (the signal transmitted by the relay is ${y_r} = {{\sqrt {{P_r}} {\kern 1pt} {x_r}} \mathord{\left/{\vphantom {{\sqrt {{P_r}} {\kern 1pt} {y_r}} {\sqrt {{P_{r,in}}} }}} \right.\kern-\nulldelimiterspace} {\sqrt {{P_{r,in}}} }}$). The signals at receiver $i$ before and after the diversity combining unit are ${{\mathbf{x}}_i'}{\kern 1pt}  = {{\mathbf{g}}_i}{y_r} + {{\mathbf{n}}_i}$ and ${x_i} = {\mathbf{c}}_i^H{{\mathbf{x}}_i'}$, respectively, where ${{\mathbf{n}}_i} \sim \mathcal{C}\mathcal{N}\left( {{{\mathbf{0}}_{{L_R} \times 1}},\sigma _i^2{\kern 1pt} {\kern 1pt} {\mathbf{I}}{}_{{L_R}}} \right)$ is the receiver thermal noise (${{\mathbf{0}}_{{L_R} \times 1}}$ is the ${L_R} \times 1$ zero vector, and ${\kern 1pt} {\mathbf{I}}{}_{{L_R}}$ is the ${L_R} \times {L_R}$ identity matrix). Finally, the SINR takes the form in \eqref{SINR} with ${\omega _{i,i}} = {\left| {{\mathbf{c}}_i^H{{\mathbf{g}}_i}{{\mathbf{h}}_i}{{\mathbf{b}}_i}} \right|^2}$, ${\omega _{j,i}} = {\left| {{\mathbf{c}}_i^H{{\mathbf{g}}_i}{{\mathbf{h}}_j}{{\mathbf{b}}_j}} \right|^2} + {{\sigma _i^2{{\left\| {{{\mathbf{c}}_i}} \right\|}^2}{{\left| {{{\mathbf{h}}_j}{{\mathbf{b}}_j}} \right|}^2}} \mathord{\left/
 {\vphantom {{\sigma _i^2{{\left\| {{{\mathbf{c}}_i}} \right\|}^2}{{\left| {{{\mathbf{h}}_j}{{\mathbf{b}}_j}} \right|}^2}} {{P_r}}}} \right.\kern-\nulldelimiterspace} {{P_r}}}$ ($j \ne i$), ${\phi _i} = {{\sigma _i^2{{\left\| {{{\mathbf{c}}_i}} \right\|}^2}{{\left| {{{\mathbf{h}}_i}{{\mathbf{b}}_i}} \right|}^2}} \mathord{\left/{\vphantom {{\sigma _i^2{{\left\| {{{\mathbf{c}}_i}} \right\|}^2}{{\left| {{{\mathbf{h}}_i}{{\mathbf{b}}_i}} \right|}^2}} {{P_r}}}} \right.\kern-\nulldelimiterspace} {{P_r}}}$, and ${\mathcal{N}_i} = \left( {{\kern 1pt} {\kern 1pt} {{\left| {{\mathbf{c}}_i^H{{\mathbf{g}}_i}} \right|}^2} + {{\sigma _i^2{{\left\| {{{\mathbf{c}}_i}} \right\|}^2}} \mathord{\left/{\vphantom {{\sigma _i^2{{\left\| {{{\mathbf{c}}_i}} \right\|}^2}} {{P_r}}}} \right.\kern-\nulldelimiterspace} {{P_r}}}} \right)\sigma _r^2$. 

As concerns the simulation parameters, we set $N = 5$, ${L_T} = {L_R} = 2$, ${P_r} = 30\;\text{dBm}$, $\varepsilon  = {10^{ - 4}}$, carrier frequency 2 GHz,  $B = 2\;\text{MHz}$, $\sigma _i^2 = \sigma _r^2 = \;F{\mathcal{N}_0}B$ (with noise figure $F = 3\;\text{dB}$ and power spectral density ${\mathcal{N}_0} =  - 174\;\text{dBm/Hz}$), ${\mu _i} = \mu  = 5$, $P_{{\kern 1pt} i}^{\max } = {P_{\max }}$, ${P_{st,i}} = {P_{st}} = 375\;\text{mW}$, and ${w_i} = {1 \mathord{\left/{\vphantom {1 N}} \right.\kern-\nulldelimiterspace} N}$ for $1 \leqslant i \leqslant N$. The distance of each transmitter/receiver from the relay is uniformly distributed in the interval [200,300] m. A path loss model with reference distance 100 m, path-loss-exponent 3.5, and standard deviation of log-normal shadowing 8 dB has been used, assuming Rayleigh fading. In addition, the QoS requirements are set as follows: $R_i^{\min } = {r_i}{\bar R_i}$, where ${r_i} \geqslant 0$ (for simplicity, ${r_i} = r$ for $1 \leqslant i \leqslant N$), and ${\bar R_i} = B\,{\log _2}\left( {1 + {{\bar \gamma }_i}} \right)$ with ${\bar \gamma _i} = {\left. {{\gamma _i}(p{{\mathbf{1}}_{N \times 1}})} \right|_{{\mathcal{N}_i} = \,0}} = {{{\omega _{i,i}}} \mathord{\left/{\vphantom {{{\omega _{i,i}}} {\left( {\sum\nolimits_{j \ne i} {{\omega _{j,i}}}  + {\phi _i}} \right)}}} \right.\kern-\nulldelimiterspace} {\left( {\sum\nolimits_{j \ne i} {{\omega _{j,i}}}  + {\phi _i}} \right)}}$ the SINR of user $i$ when all the transmit powers are equal and the equivalent noise power is zero. Unless otherwise stated, the initial point is selected as ${\mathbf{p}} = {P_{\max }}{{\mathbf{1}}_{N \times 1}}$ (we assume $0 \leqslant r < 1$, since this point is infeasible when $r \geqslant 1$). All the results are derived from the statistical average of ${10^4}$ independent problem instances.

First of all, we examine the convergence speed of Algorithm 1 through numerical analysis, since it is difficult to be studied analytically. Fig. 1 shows that Algorithm 1 always generates an increasing sequence and converges very fast within only a few iterations. Thus, Algorithm 1 exhibits low complexity because the number of iterations until convergence is quite small and the convex problem in each iteration can be globally solved in polynomial time using standard convex optimization techniques, such as interior-point methods [8]. Furthermore, Algorithm 1 is robust since different initialization points may achieve slightly different final objective values, and also the convergence speed remains almost the same. 

\begin{figure}[!t]
\centering
\includegraphics[width=3.5in]{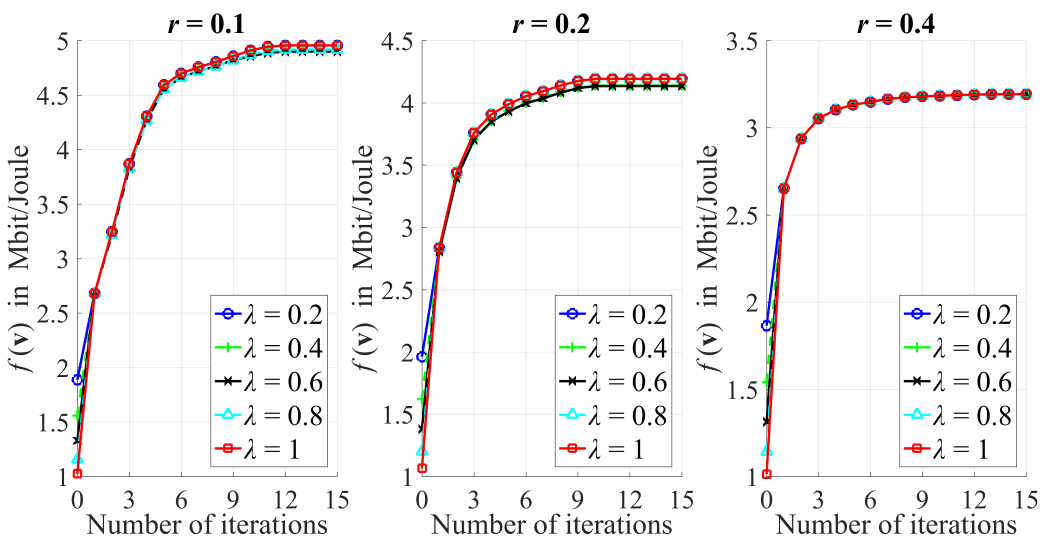} % max width=3.5in
\caption{Convergence of Algorithm 1 (WSEE maximization), with ${P_{\max }} = 20\;\text{dBm}$, for different QoS requirements and initial point ${\mathbf{p}} = \lambda {P_{\max }}{{\mathbf{1}}_{N \times 1}}$.}
\label{Fig1}
\end{figure}

Subsequently, for the sake of comparison, we introduce a baseline scheme, namely, weighted-sum rate (WSR) maximization defined as follows:
\begin{equation}
\mathop {\max }\limits_{{\mathbf{p}}{\kern 1pt}  \in {\kern 1pt} S} \quad \text{WSR}({\mathbf{p}}) = \sum\nolimits_{i = 1}^N {{w_i}{R_i}({\mathbf{p}})}
\end{equation}
This problem is solved by SCO, using again the transformation ${\mathbf{p}} = {2^{\mathbf{q}}}$, where the convex problems take the form:
\begin{equation}
\mathop {\max }\limits_{{\mathbf{q}}{\kern 1pt}  \in {\kern 1pt} \Theta } \quad \sum\nolimits_{i = 1}^N {{w_i}\widetilde {R_i'}({\mathbf{q}})}
\end{equation}
with feasible set $\Theta  = \{ {\mathbf{q}} \in {\mathbb{R}^N}:{\kern 1pt} \;{2^{{q_i}}} \leqslant P_{{\kern 1pt} i}^{\max }\;\text{and}\allowbreak\;{\vartheta _i}({\mathbf{q}}) \geqslant 0,\;1 \leqslant i \leqslant N\}$. Figs. 2 and 3 illustrate respectively the achieved WSEE and WSR versus ${P_{\max }}$ for different QoS requirements. In Fig. 2, we can observe that: 1) for each scheme, the increase of QoS requirements leads to the decrease of WSEE because the feasible set becomes smaller, and 2) for low ${P_{\max }}$, WSEE and WSR maximization are almost equivalent, since $\text{WSEE}({\mathbf{p}}) \approx ({1 \mathord{\left/{\vphantom {1 {{P_{st}}}}} \right.\kern-\nulldelimiterspace} {{P_{st}}}})\text{WSR}({\mathbf{p}})$ ($\mu {\kern 1pt} {p_i} \leqslant \mu {\kern 1pt} {P_{\max }} \ll {P_{st}}\; \Rightarrow \;{P_{c,i}}({p_i}) \approx {P_{st}}$), while WSEE increases with ${P_{\max }}$. Similar observations can be made in Fig. 3. Nevertheless, for larger values of ${P_{\max }}$, it can be seen that: 1) in Fig. 2, WSEE remains constant when maximizing the WSEE, whereas decreases with ${P_{\max }}$ when maximizing the WSR because of the higher required transmit power, and 2) in Fig. 3, WSR maximization achieves slightly higher WSR than WSEE maximization, while both schemes reach a peak value (note that WSR is upper-bounded when ${\phi _i} \ne 0$: $\text{WSR}({\mathbf{p}}) \leqslant \sum\nolimits_{i = 1}^N {{w_i}B\,{{\log }_2}\left( {1 + \gamma _i^{\max }} \right)} $ with $\gamma _i^{\max } = \mathop {\lim }\limits_{{p_i} \to \infty } {\gamma _i}\left( {\mathbf{p}} \right) = {{{\omega _{i,i}}}\mathord{\left/{\vphantom {{{\omega _{i,i}}} {{\phi _i}}}} \right.\kern-\nulldelimiterspace} {{\phi _i}}}$). 

\begin{figure}[!t]
\centering
\includegraphics[width=3.5in]{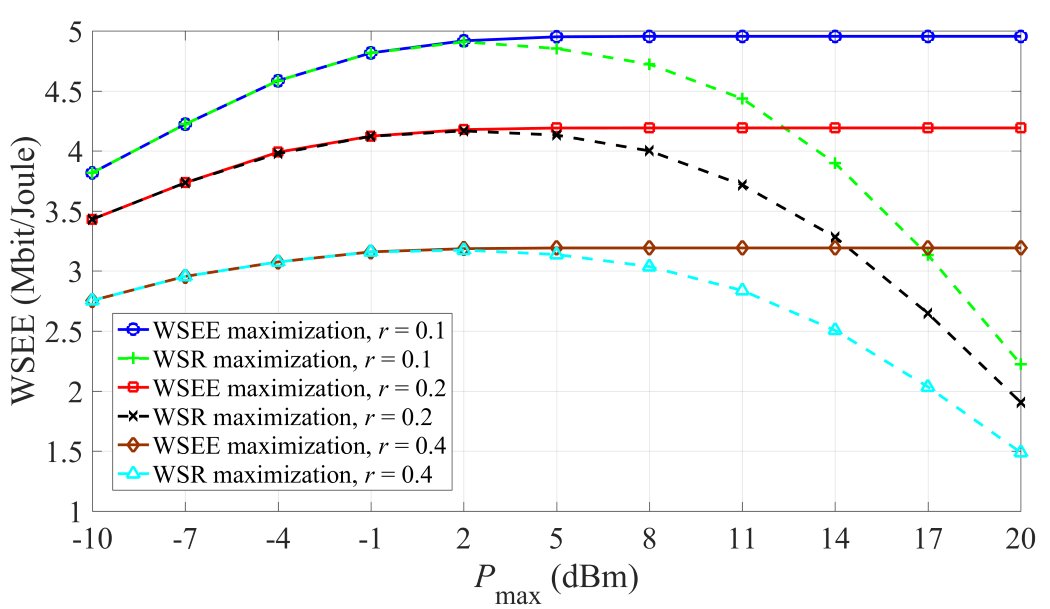} % max width=3.5in
\caption{Achieved WSEE versus ${P_{\max }}$ by maximizing: a) the WSEE (Algorithm 1), and b) the WSR (baseline scheme) for different QoS requirements.}
\label{Fig2}
\end{figure}

\begin{figure}[!t]
\centering
\includegraphics[width=3.5in]{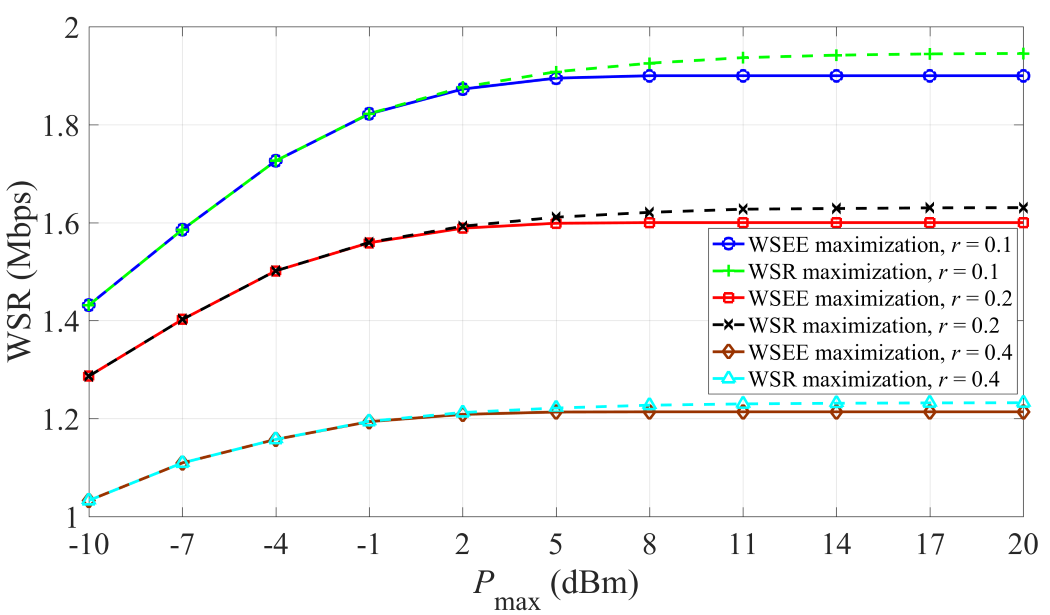} % max width=3.5in
\caption{Achieved WSR versus ${P_{\max }}$ by maximizing: a) the WSEE (Algorithm 1), and b) the WSR (baseline scheme) for different QoS requirements.}
\label{Fig3}
\end{figure}

\section{Conclusion}
In this paper, we have presented a general methodology for WSEE maximization in wireless networks. More specifically, we have developed a low-complexity and robust algorithm that is theoretically guaranteed to converge and is able to achieve a KKT solution. Finally, we have studied notable extensions of the proposed approach to systems with multiple resource blocks and general power consumption model as well.

%\clearpage
%\balance

\end{document}